\documentstyle[preprint,aps,epsf,amstex]{revtex}

\def\ORDER#1{\hbox{${\cal O}(#1)$}}
\def\NOBAR#1{#1}
\def\BAR#1{\overline{#1}}

\def\KET#1{\vert #1 \rangle}
\def\BRACKET#1#2{\langle #1 \vert #2 \rangle}

\begin{document}
\draft
\tighten

\title{On the Problem of Programming Quantum Computers}

\author{Hans De Raedt, Anthony Hams, and Kristel Michielsen}

\address{
Institute for Theoretical Physics and Materials Science Centre \\
University of Groningen, Nijenborgh 4 \\
NL-9747 AG Groningen, The Netherlands \\
E-mail: deraedt@@phys.rug.nl, A.H.Hams@@phys.rug.nl, kristel@@phys.rug.nl \\
http://rugth30.phys.rug.nl/compphys
}

\author{Seiji Miyashita and Keiji Saito}

\address{
Department of Applied Physics, School of Engineering \\
University of Tokyo, Bunkyo-ku, Tokyo 113, Japan \\
E-mail: miya@@yuragi.t.u-tokyo.ac.jp, saitoh@@spin.t.u-tokyo.ac.jp
}

\date{\today}
\maketitle
\begin{abstract}
We study effects of the physical realization of quantum computers
on their logical operation.
Through simulation of physical models of quantum computer hardware,
we analyse the difficulties that are encountered in programming
physical implementations of quantum computers. We discuss the origin
of the instabilities of quantum algorithms and explore physical mechanisms
to enlarge the region(s) of stable operation.
\end{abstract}

\newpage

\section{Introduction}
The logical operation of a conventional digital computer
does not depend on the details of the hardware implementation
although the speed of operation and the cost of the machine obviously do.
Conventional computers are in one particular state at any time.
Furthermore from the point of view of programming the computer,
the internal machinery of the basic units comprising the computer is irrelevant.
This is very important because it implies that on a conceptual level,
algorithms designed to run on a conventional computer will produce
answers that do not depend on the hardware that is used.

A quantum computer (QC) is very different in this respect.
A QC exploits the fact that a quantum system can be
in a superposition of states
and that interference of these states allows exponentially
many computations to be done in parallel
\cite{Feynman:1982,DiVincenzo:1995,Ekert:1996,Vedral:1998,Aharonov:1998}.
The presence of this superposition is a manifestation
of the internal quantum dynamics of the elementary units
(i.e. the qubits).
In other words, the quantum dynamics is essential
to the operation of a physically realizable QC.

The operation of an ideal QC does not depend on
the intrinsic dynamics of the physical qubits:
One imagines that the qubits are ideal two-state
quantum systems that perform their task instantaneously and
perfectly. From a theoretical point of view
this situation is very similar to that of computers built from
conventional digital circuits.
However, in practice there is a fundamental difference.
The fact that the logical operation of conventional digital circuits
does not depend on their hardware implementation
(e.g. semiconductors, relays, vacuum tubes, etc.) is
directly related to the presence of dissipative processes
that drive the circuits into regions of stable operation.
Dissipation suppresses the effects of the internal, non-ideal
(chaotic) dynamics of these circuits.

The quantum dynamics of small physical devices is usually very
sensitive to small perturbations and this holds for qubits as well.
Unfortunately, in contrast to the case of conventional circuits,
dissipation usually has a devastating effect on the coherent quantum
dynamical motion of the qubits, i.e. on the very essence of QC's.
Therefore the specific physical realization of a QC is intimately related to
the stability of its operation.

In this paper we study the relation
between the physical realization of QC's and their logical operation
and explore physical mechanisms to enlarge the region(s) of
stable operation.
We demonstrate that programming a physical implementation of a QC
is non-trivial, even if the QC consists of only two or four qubits.
In most theoretical work on QC's and quantum algorithms (QA's)
\cite{Aharonov:1998,Ekert:1998,Shor:1994prime,Chuang:1995,Grover:1996,Grover:1997prl,Grover:1998,Shor:1999,Kim:2000,Gingrinch:2000}
one considers theoretically ideal (but physically unrealizable) QC's
and therefore this problem of programming QC's
(which we will call the Quantum Programming Problem (QPP)) is not an issue.
As far as we know no experimental data has been published
that specifically addresses this, for potential applications,
very important and intrinsic problem of programming QC's.
The aim of this paper is to investigate various aspects of the
QPP by simulating QC hardware.

How does a QPP reveal itself?
Consider two logically independent operations ($O_1$ and $O_2$) of the machine.
On a conventional computer or ideal QC, the order
in which we execute these two mutually independent
instructions does not matter: $O_1 O_2 = O_2 O_1$.
However, it turns out that
on a physically realizable QC {\sl sometimes} the order does matter, even
if there are no dependencies in these two program steps.
In some cases $O_1 O_2 \not= O_2 O_1$ and the QC may
give the wrong answers.
The QPP is due to the fact that
we are dealing with interacting quantum mechanical objects
(as communication between qubits is essential for computation),
technical difficulties to manipulate a qubit without disturbing
others and the fundamental physical fact that the state of a qubit
cannot be frozen during the time that other qubits are being
addressed.
Also note the qualifier {\sl sometimes}. There seems to be
no general rule to decide beforehand which operation
and at what stage of the QA the QPP leads to incorrect results.
At present the only way
to find out seems to be to actually carry out the calculations and
check the results.

It does not require a lot of imagination to see that the QPP implies
that it may be very difficult to develop a non-trivial quantum program
for a physical QC.
Moreover there is no guarantee that a QA that works
well on one QC will perform well on other QC's. Marginal
changes in the qubit hardware may affect the interchangeability
of logically independent operations.
There are several factors that contribute to the QPP:
%
\begin{itemize}
\item[1)]
Differences between the theoretically perfect
and physically realizable one- and two-qubit operations, e.g.
the one-qubit operations affecting other qubits and inaccuracies
on the time-interval used to perform operations.
\item[2)]
Physical qubits cannot be kept still while others
are being addressed.
\item[3)]
The effect of coupling of the qubits to
other degrees of freedom (dissipation, decoherence).
\end{itemize}

In this paper we address these issues through case studies.
In Section~\ref{sec:physmod} we describe the physical model that will
be the starting point of our investigations.
Our choice of physical models is largely inspired by NMR-QC
experiments
\cite{Jones:1998,Jones:1998nat,Chuang:1998nat,Chuang:1998prl,Marx:1999,Knill:2000},
only because other candidate technologies
\cite{Cirac:1995,Monroe:1995,Sleator:1995,Domokos:1995,Kane:1998,Imamoglo:1999,Makhlin:1999,Nakamura:1999,Nogues:1999,Molmer:1999,Sorensen:1999,Fazio:1999,Orlando:1999,Blias:2000,Oliveira:2000}
for building QC's are not yet developed to the point that they can execute
computationally non-trivial QA's.

As an example of such a QA we will
take Grover's database search algorithm (see Section~\ref{sec:grov})
and implement it on various physical models for 2- and 4-qubit QC's
(Sections~\ref{sec:twoqubits} to~\ref{sec:fourqubits}).
Our approach for analysing the QPP is
to run Grover's QA by simulating the time-evolution of
the physical model representing the QC. Thereby we
strictly stick to the rules of quantum mechanics, i.e.
we solve the time-dependent Schr\"odinger equation that
describes the evolution of the physical apparatus representing
the QC. The main vehicle for performing these simulations is
a Quantum Computer Emulator
(QCE)\cite{QCEdownload}.
A detailed description of this software tool is given elsewhere
\cite{DeRaedt:2000qce,DeRaedt:2000sdqc}.
Our work is fundamentally different from those of
others who also address questions related to error propagation
in QA's
\cite{Miquel:1995,Pablo:2000,Long:2000,Berman:2000sim}
in that we execute QA's on physical models of QC hardware.
The influence of non-resonant effects on the quantum computations
using Ising spin quantum computers has been studied by Berman et al.
\cite{Berman:2000nr}. This work is similar in spirit to the one of
the present paper as it explores the consequences of the difference
between the ideal and physically realizable QC's.
In this paper we focus on the QPP, not on methods to
suppress non-resonant effects.
For simplicity most of our calculations (Sections~\ref{sec:twoqubits} and~\ref{sec:fourqubits})
are done for systems at zero temperature,
in the absence of interactions with the environment.
Simulations of QC's
operating at a non-zero temperature, in contact with a heat bath,
are discussed in Section~\ref{sec:dissip}.

\section{Physical Model}
\label{sec:physmod}
The simplest qubit is a two-state quantum system,
e.g. the spin of electrons or the polarization of photons.
The basic operations in a meaningful computation are the manipulation
of each qubit (e.g. by applying external fields)
and the exchange of information between the qubits.
In physical terms, the latter implies
that there should be some interaction between the qubits.
A non-trivial QC contains at least two qubits.
It is known that the most simple spin-1/2 system,
i.e. the Ising model, can be used for quantum computing
\cite{Lloyd:1993,Berman:1994,Berman:1998qc}.
In the presence of an external magnetic field, the Hamiltonian
of the two-spin Ising model reads
\begin{align}
\label{eq:HAM}
H =- J_{z} S_1^z S_2^z - h_{z} (g_1 S_1^z+ g_2 S_2^z)
,
\end{align}
where $J_z$, $g_1$ and $g_2$ are material-specific constants,
$h_z$ represents the applied magnetic field,
and $S_j^\alpha$ denotes the $\alpha$-th component
($\alpha=x,y,z$) of the spin-1/2 operators describing the nuclear
spins. In this paper we use units such that $\hbar=1$.

According to the rules of quantum mechanics,
the state of the QC at time $t$ is completely described by
the wave function $\KET{\Phi(t)}$.
Executing a program on a QC is equivalent to solving
the time-dependent Schr\"odinger equation (TDSE)
\begin{align}
\label{eq:TDSE}
i
{\partial \over\partial t} \KET{\Phi(t)}= H(t) \KET{\Phi(t)}
.
\end{align}
For the 2-qubit QC the most general linear combination reads
\begin{align}
\KET{\Phi(t)}=
a(\downarrow,\downarrow;t)
\KET{\downarrow\downarrow}
+a(\uparrow,\downarrow;t)
\KET{\uparrow\downarrow}
+a(\downarrow,\uparrow;t)
\KET{\downarrow\uparrow}
+a(\uparrow,\uparrow;t)
\KET{\uparrow\uparrow}
,
\end{align}
where the $a(b_1,b_2;t)$ are complex numbers.
The probability that, upon measurement,
the QC is in one of the four basis states
$\KET{\downarrow\downarrow}, \ldots, \KET{\uparrow\uparrow}$
is given by
$|a(\downarrow,\downarrow;t)|^2,\ldots,|a(\uparrow,\uparrow;t)|^2$
respectively.

The final results of a QC calculation can be read off
by performing an experiment that measures the expectation
value(s) of the spin(s).
The value of a qubit is related to
the expectation value of the $z$-component of the spin operator:
\begin{align}Q_j\equiv 1/2 -\BRACKET{\Phi(t)}{S^z_j \vert \Phi(t)}\quad;\quad
j=1,2,\ldots,N,
\end{align}
where $N$ denotes the total number of qubits of the QC.
In this paper we will denote the state of a qubit by
\begin{align}
\KET{0}\equiv\KET{\uparrow}
\quad;\quad
\KET{1}\equiv\KET{\downarrow}
,
\end{align}
and if necessary we will add a subscript, e.g. $\KET{0}_1$, to label the
qubit.

The Ising interaction between the spins is sufficient to
implement control-NOT (CNOT) gates. Single qubit
operations can be performed by applying additional
external fields in the $x$ and $y$ direction (see below).
It has been shown that any operation involving an arbitrary
number of qubits can be written
as a sequence of these elementary operations\cite{Barenco:1995},
in other words the Ising model is a universal QC.

\section{Grover's database search algorithm}
\label{sec:grov}
Finding the needle in a haystack of $N$ elements requires \ORDER{N}
queries on a conventional computer\cite{Cormen:1994}.
Grover has shown that a QC can find the needle using
only \ORDER{\sqrt{N}} attempts\cite{Grover:1996,Grover:1997prl}.
Assuming a uniform probability distribution for the needle,
for $N=4$ the average number of queries required by a conventional
algorithm is 9/4\cite{Chuang:1998prl,Cormen:1994}.
With Grover's QA the correct answer can be found in
a single query\cite{Jones:1998nat,Chuang:1998prl}.
In general the reduction from \ORDER{N} to \ORDER{\sqrt{N}} is due to
the intrinsic massive parallel operation of the QC.

Experimentally this QA has been implemented on a 2-qubit NMR-QC for
the case of a database containing four items\cite{Jones:1998nat,Chuang:1998prl}.
This implementation uses
elementary rotations about 90 degrees (clock-wise)
around the $x$ and $y$-axis
(e.g. ${\NOBAR X_1}\KET{00}=(\KET{00}+i\KET{10})/\sqrt{2}$
etc.)
and an interaction-controlled phase shift\cite{Jones:1998nat,Chuang:1998prl}.
It is convenient to express these procedures in matrix notation, for example
\begin{align}
\label{eq:X1}
X_1
\left(
\begin{array}{l}
\KET{00}\\ \KET{10}\\ \KET{01}\\ \KET{11}
\end{array}
\right)
\equiv{1\over\sqrt{2}}
\left(
\begin{array}{rrrr}
1&i&0&0 \\
i&1&0&0 \\
0&0&1&i \\
0&0&i&1
\end{array}
\right)
\left(
\begin{array}{l}
\KET{00}\\ \KET{10}\\ \KET{01}\\ \KET{11}
\end{array}
\right)
,
\end{align}
where $\KET{b_1 b_2}\equiv\KET{b_1}\KET{b_2}$ and $b_i=0,1$.
\begin{align}
\label{eq:Y2}
Y_2
\left(
\begin{array}{l}
\KET{00}\\ \KET{10}\\ \KET{01}\\ \KET{11}
\end{array}
\right)
\equiv{1\over\sqrt{2}}
\left(
\begin{array}{rrrr}
1&0&1&0 \\
0&1&0&1 \\
-1&0&\phantom{-}1&\phantom{-}0 \\
0&-1&0&1
\end{array}
\right)
\left(
\begin{array}{l}
\KET{00}\\ \KET{10}\\ \KET{01}\\ \KET{11}
\end{array}
\right)
.
\end{align}
The two qubits ``communicate'' with each other through
the interaction-controlled phase shift
\begin{align}
\label{eq:Ia}
I(a)
\left(
\begin{array}{l}
\KET{00}\\ \KET{10}\\ \KET{01}\\ \KET{11}
\end{array}
\right)
\equiv{1\over\sqrt{2}}
\left(
\begin{array}{rrrr}
e^{-ia/4}&0&0&0 \\
0&e^{+ia/4}&0&0 \\
0&0&e^{+ia/4}&0 \\
0&0&0&e^{-ia/4}
\end{array}
\right)
\left(
\begin{array}{l}
\KET{00}\\ \KET{10}\\ \KET{01}\\ \KET{11}
\end{array}
\right)
.
\end{align}
This unitary transformation can be used to implement a
CNOT gate. Grover's algorithm can be expressed entirely by a sequence
of the basic operations (\ref{eq:X1}),(\ref{eq:Y2}), and (\ref{eq:Ia}).

To see how Grover's QA works it is instructive to consider an example
of a database with $N=4$ positions, labeled 0,...,3.
Let us assume that the item we are searching for is located at position 2.
First we put the QC in its initial state $\KET{00}$.
Then we transform this state into the uniform superposition state
\begin{align}
\KET{u}\equiv{1\over 2}(\KET{00}+\KET{10}+\KET{01}+\KET{11})
,
\end{align}
by letting the sequence
${\BAR X}_2 {\BAR X}_2 {\BAR Y}_2{\BAR X}_1 {\BAR X}_1 {\BAR Y}_1$
act on $\KET{00}$\cite{Chuang:1998prl}:
\begin{align}
\label{eq:u1}
\KET{u}=
{\BAR X}_2 {\BAR X}_2 {\BAR Y}_2
{\BAR X}_1 {\BAR X}_1 {\BAR Y}_1
\KET{00}
,
\end{align}
where $\BAR X_j$ ($\BAR Y_j$) denotes the inverse of $X_j$ ($Y_j$).

Next we apply to $\KET{u}$ the sequence of elementary
operations\cite{Jones:1998nat,Chuang:1998prl,DeRaedt:2000qce,DeRaedt:2000sdqc}
\begin{align} \label{eq:f2}
F_2\equiv
{\BAR Y}_1 {\NOBAR X}_1 {\BAR Y}_1
{\NOBAR Y}_2 {\BAR X}_2 {\BAR Y}_2
I(\pi)
,
\end{align}
to encode the content of the database as
\begin{align}
\label{eq:content}
\KET{\Psi_2}={1\over 2}(\KET{00}+\KET{10}-\KET{01}+\KET{11})
,
\end{align}
where we adopt the notation in which the basis states are labeled by
the binary representation of integers with the order of the bits
reversed.
In~(\ref{eq:content}) the position of the item in the database (i.e. 2 in this example)
is encoded by modifying $\KET{u}$ such that
the amplitude of the corresponding basis state changes sign.

The key ingredient of Grover's algorithm is
an operation that determines which of the basis state
contributes to~(\ref{eq:content}) with the minus-one amplitude.
In matrix notation
\begin{align} \label{eq:Gmat}
G={1\over2}
\left(
  \begin{array}{rrrr}
    -1 & 1 & 1 & 1 \\
    1 & -1 & 1 & 1 \\
    1 & 1 & -1 & 1 \\
    1 & 1 & 1 & -1
  \end{array}
\right)
,
\end{align}
and in terms of elementary operations
\begin{align}
\label{eq:G}
G=
{\NOBAR X}_1 {\NOBAR X}_1 {\BAR Y}_1
{\NOBAR X}_2 {\NOBAR X}_2 {\BAR Y}_2
{\NOBAR Y}_1 {\BAR X}_1 {\BAR Y}_1 {\NOBAR Y}_2 {\BAR X}_2 {\BAR Y}_2 I(\pi)
{\NOBAR X}_1 {\NOBAR X}_1 {\BAR Y}_1
{\NOBAR X}_2 {\NOBAR X}_2 {\BAR Y}_2
.
\end{align}

The sequence~(\ref{eq:G}) is by no means unique: Various alternative
expressions can be written down by using the algebraic
properties of the $X$'s and $Y$'s.
This feature can be exploited to eliminate redundant
elementary operations\cite{Chuang:1998prl}.
Starting from the uniform superposition state $\KET{u}$,
one choice for the optimized sequences that implement
the four different states of the database
and Grover's search algorithm is\cite{Jones:1998nat,Chuang:1998prl,DeRaedt:2000qce,DeRaedt:2000sdqc}
\begin{subequations} 
\label{eq:sequence}
\begin{align} 
U_0 &={     X}_1{\BAR Y}_1 {     X}_2{\BAR Y}_2 I(\pi)
     {     X}_1 {\BAR Y}_1 {     X}_2 {\BAR Y}_2 I(\pi), \\
U_1 &={     X}_1{\BAR Y}_1 {     X}_2{\BAR Y}_2 I(\pi)
{     X}_1 {\BAR Y}_1 {\BAR X}_2 {\BAR Y}_2 I(\pi),  \\
U_2 &={     X}_1{\BAR Y}_1 {     X}_2{\BAR Y}_2 I(\pi)
{\BAR X}_1 {\BAR Y}_1 {     X}_2 {\BAR Y}_2 I(\pi), \\
U_3 &={     X}_1{\BAR Y}_1 {     X}_2{\BAR Y}_2 I(\pi)
{\BAR X}_1 {\BAR Y}_1 {\BAR X}_2 {\BAR Y}_2 I(\pi),
\end{align}
\end{subequations} 
where the $U_n$ correspond to the case where the needle is in position $n$.

The sequences of unitary transformations, e.g. $X_1,Y_2$,etc.,
are easily emulated on an ordinary computer.
We use our software package, Quantum Computer Emulator (QCE) \cite{DeRaedt:2000qce},
to perform these calculations.
An example will be shown in Appendix~\ref{app:qce}.
On an ideal QC, sequences~(\ref{eq:sequence}) return the correct answer, i.e.
the position of the searched-for item.
This is easily verified on the QCE
by selecting the elementary operations (called micro instructions
on the QCE) that implement an ideal QC.

\section{Two-qubit QC's}
\label{sec:twoqubits}
The energy-level structure of
the nuclear spins of molecules such as deuterated cytosin\cite{Jones:1998,Jones:1998nat,Jones:1999} and
carbon-13 labeled chloroform\cite{Chuang:1998nat,Chuang:1998prl}
can be described by model~(\ref{eq:HAM}) and hence they can
be used as physical realizations of 2-qubit QC's.

\subsection{Resonant pulses}
\label{sec:respulses}
NMR-QC experiments on carbon-13 labeled chloroform\cite{Chuang:1998nat}
use resonant pulses to manipulate the quantum state of
the nuclear spins of the $^1$H and $^{13}$C atoms.
In the presence of a static magnetic field along the $+z$ direction
this NMR-QC system is described by~(\ref{eq:HAM})
with $h_{z}g_1/2\pi\approx 500 \hbox{MHz}$,
$h_{z}g_2/2\pi\approx125 \hbox{MHz}$, and
$J_{z}/2\pi\approx-215 \hbox{Hz}$\cite{Chuang:1998nat}.
A detailed account of simulations for this case
have been published elsewhere\cite{DeRaedt:2000qce,DeRaedt:2000sdqc}.
Simulations for physical model~(\ref{eq:HAM}) confirm
that sequences~(\ref{eq:sequence}) yield the correct answers
for the database search problem\cite{DeRaedt:2000qce}.
However, we also demonstrated that the outcome of these
calculations may be very sensitive to the order in which
logically independent operations are carried out\cite{DeRaedt:2000sdqc}..
Some of these results are reproduced in Table~\ref{tab:grover} (first six rows,
for details see \cite{DeRaedt:2000qce,DeRaedt:2000sdqc}).

The results marked with a tilde are obtained by using
a logically identical but physically different uniform
superposition, i.e.
\begin{align} \label{eq:uprime}
\KET{u^\prime}=
{\BAR X}_1 {\BAR X}_1 {\BAR Y}_1
{\BAR X}_2 {\BAR X}_2 {\BAR Y}_2
\KET{00}
.
\end{align}
On an ideal QC, $\KET{u^\prime}=\KET{u}$ but on a physical realizable
machine this is unlikely to be the case.
In an experiment it is simply impossible to freeze spin 2 (1)
during the time that resonant pulses are being applied to spin 1 (2).
Unless the length of these pulses is chosen judiciously,
the wave function will acquire an additional phase.
The corresponding unitary transformation does not necessarily
commute with the operations that follow, potentially leading
to an incorrect final result (as shown in Table~\ref{tab:grover}), as we
pointed out in a previous paper\cite{DeRaedt:2000sdqc}.

For the two-spin system~(\ref{eq:uprime}) one may optimize
the pulse durations such that the effect of these phase errors
yields qualitatively correct answers.
A basic step is to make the pulse lengths commensurate
with all relevant time scales\cite{Berman:2000nr,Berman:1998qc}.
The main idea of the method suggested in\cite{Berman:1998qc}
and demonstrated in\cite{Berman:2000nr} is to choose the parameters
of the electromagnetic pulse such that the non-resonant spin still rotates
but returns to its initial position at the end of the pulse \cite{Berman:2000nr}.
The results of the calculations are shown in Fig.~\ref{fig:qcesnap} and
summarized in Table~\ref{tab:grover} (primed
symbols)\cite{QCEdownload2}.
In the numerical calculations we used $J=-10^{-6}$,
$h_zg_1=1$, $h_zg_2=0.25$ ,
($h_yg_1=0.025$, $h_yg_2=0.0625$, $t_{\hbox{pulse}}=4\pi$),
for the pulses on the first spin and
($h_yg_1=0.05$, $h_yg_2=0.0125$, $t_{\hbox{pulse}}=8\pi$)
for the pulses on the second spin, all values in dimensionless units.
It is clear that optimization has the desired effect
on the sequences that operate on $\KET{u^\prime}$ (symbols
with a tilde).

In spite of this optimization, Table~\ref{tab:grover} shows that there are
significant quantitative differences between the theoretically exact results
(rows (1,2)) and those obtained by simulating a physical model
of a QC (e.g. rows 7 to 10).
Even if the pulse length is taken to be commensurate
with the relevant time scales of the QC,
changing the state of the qubits
by way of resonant pulses yields quantum states
that are different from those obtained by means of
the unitary transformations used in the analysis of the ideal QC.
This is because spin 2 also interacts with the field applied
to spin 1 and vice versa.
With each program step, the non-ideal unitary operation may
or may not result in the proliferation of errors.
These errors are systematic (there is no ``random'' error source
in our calculations) and directly linked to the structure of the QA.
This is a clear case of a QPP, although we managed to let the
different QA's produce the correct answer.
Note that the QPP cannot be solved by means of error correction\cite{Kak:1999}: 
The operations on the extra qubits required for error correction
will suffer from exactly the same QPP.

\subsection{Stability of Grover's quantum algorithm}
So far we studied the stability of quantum algorithms
by perturbing the input to the database encoding part of the algorithm.
In this subsection we will study the QPP of the database query part of
Grover's search algorithm (the operation $G$, see~(\ref{eq:G})) itself.
First we will assume that the input provided to $G$ is exact
(i.e. of the form~(\ref{eq:u1}) for example) and we will compare
the output of logically identical but physically different
implementations of $G$. As examples we take the original
sequence
\begin{align}
\label{eq:G1}
G=
{\NOBAR X}_1 {\NOBAR X}_1 {\BAR Y}_1
{\NOBAR X}_2 {\NOBAR X}_2 {\BAR Y}_2
{\NOBAR Y}_1 {\BAR X}_1 {\BAR Y}_1 {\NOBAR Y}_2 {\BAR X}_2 {\BAR Y}_2 I(\pi)
{\NOBAR X}_1 {\NOBAR X}_1 {\BAR Y}_1
{\NOBAR X}_2 {\NOBAR X}_2 {\BAR Y}_2
,
\end{align}
and two, logically identical, sequences
\begin{align}
\label{eq:G2}
{\hat G}=
{\NOBAR X}_1 {\NOBAR X}_1 {\BAR Y}_1
{\NOBAR X}_2 {\NOBAR X}_2 {\BAR Y}_2
{\NOBAR Y}_1 {\BAR X}_1 {\NOBAR Y}_2 {\BAR X}_2 {\BAR Y}_1 {\BAR Y}_2
I(\pi)
{\NOBAR X}_1 {\NOBAR X}_1 {\BAR Y}_1
{\NOBAR X}_2 {\NOBAR X}_2 {\BAR Y}_2
,
\end{align}
\begin{align}
\label{eq:G3}
{\tilde G}=
{\NOBAR X}_1 {\NOBAR X}_1 {\BAR Y}_1
{\NOBAR X}_2 {\NOBAR X}_2 {\BAR Y}_2
{\NOBAR Y}_2 {\BAR X}_2 {\BAR Y}_2
{\NOBAR Y}_1 {\BAR X}_1 {\BAR Y}_1
I(\pi)
{\NOBAR X}_2 {\NOBAR X}_2 {\BAR Y}_2
{\NOBAR X}_1 {\NOBAR X}_1 {\BAR Y}_1
.
\end{align}
Note that on purpose we did not ``optimize'' these sequences by using
e.g. ${\BAR X}_1 {\NOBAR X}_1=1$.
On an ideal QC we have\cite{Chuang:1998prl}
\begin{align}
G\KET{\Psi_2}=G^3\KET{\Psi_2}=
{\hat G}\KET{\Psi_2}={\hat G}^3\KET{\Psi_2}=
{\tilde G}\KET{\Psi_2}={\tilde G}^3\KET{\Psi_2}=\KET{01}
,
\end{align}
providing another test of the stability of the query operation $G$
on a physical QC. Table~\ref{tab:G} contains the numerical results
obtained by running the sequences~(\ref{eq:G1}), (\ref{eq:G2}), and~(\ref{eq:G3})
on the QCE using the exact state $\KET{\Psi_2}$ (see~(\ref{eq:G})) as input.
In the case of the NMR-like QC optimized resonant pulses were used.
The ideal QC performs as expected but the physical implementation
(symbols with a prime) does not.
In fact even one application of ${\tilde G}$ apparently
returns an answer that is close to being useless ($Q^\prime_1\approx0.5$).
As the three sequences ~(\ref{eq:G1}), (\ref{eq:G2}), and~(\ref{eq:G3}) are logically identical this
is a clear case of a QPP.

The occurrence of a QPP seems to be a generic feature of QA's running
on QC's. Therefore it is of interest to try to quantify the QPP.
We now describe a simple procedure for this purpose, using $G$
and the case where the item is located in position 0 (the exact input
state being $\KET{\Psi_0}$) as an example.
In general there are two sources of errors in this calculation:
The input $\KET{\Psi^\prime}$ to $G$ and $G$
itself, the latter depending on the particular hardware implementation
of the QC. As before we will use the resonant pulse technique in our numerical
experiments.

We write $\KET{\Psi^\prime}$ as
\begin{align}
\KET{\Psi^\prime}=
\alpha_0\KET{\Psi_0}
+\alpha_1\KET{\Psi_1}
+\alpha_2\KET{\Psi_2}
+\alpha_3\KET{\Psi_3}
,
\end{align}
where the amplitude $\alpha_0$ can always be taken real
($-1\le \alpha_0\le 1$) and is chosen at random.
The other three complex coeffients are chosen randomly too,
subject to the constraint
$|\alpha_1|^2+|\alpha_2|^2+|\alpha_3|^2=1-|\alpha_0|^2$.
The real variable $x\equiv\BRACKET{\Psi^\prime}{\Psi_0}=\alpha_0$
is a measure for how much the input state
deviates from the exact reference input $\KET{\Psi_0}$.
On an ideal QC, $G\KET{\Psi_0}=\KET{00}$. Thus we can use
the state $\KET{00}$ as reference to determine how much
the output state $\KET{\Theta}\equiv G\KET{\Psi^\prime}$
deviates from the exact answer.
We quantify this deviation by the variable
$y=|\BRACKET{00}{G\Psi^\prime}|$.

\begin{figure}[t]
\def\epsfsize#1#2{1.2#1} 
\begin{center}
\epsffile{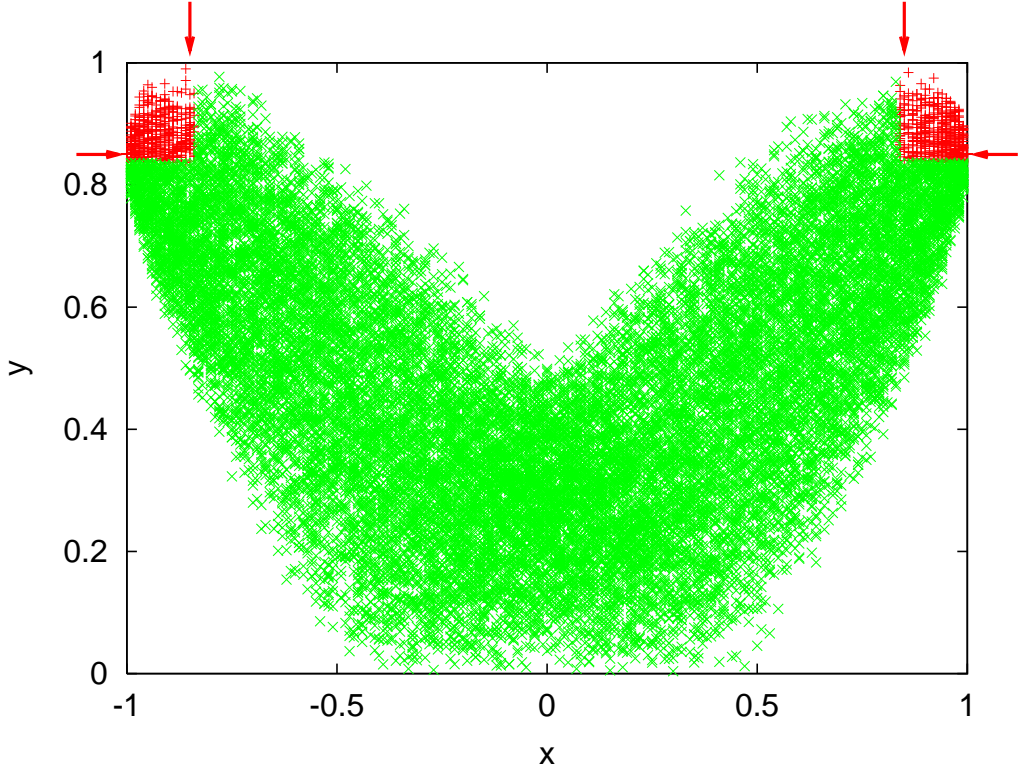}
\end{center}
\caption{Region of stable (black) and unstable (gray) operation of Grover's
search algorithm when executed on a 2-qubit NMR-like QC.
The $x$-coordinate ($x=\BRACKET{\Psi^\prime}{\Psi_0}$)
gives the projection of the random
input state $\KET{\Psi^\prime}$ on the exact input state $\Psi_0$, i.e.
for the case where the item is at position 0.
The $y$-coordinate ($y=|\BRACKET{00}{G\Psi^\prime}|$)
discriminates between correct and false output.
A confidence level of $c=0.7$ on the input-output pair
was used to determine if $G$ yields the correct result.
The horizontal (vertical) arrows mark the values of $y$ ($x$)
at which the input (output) state is equal to its ideal value.
Note the absence of points in the regions
near $(-1,1)$ and $(1,1)$.}
\label{fig:scatter}
\end{figure}

The result of a numerical experiment using 20000 random input states
$\KET{\Psi_0}$ is shown in Fig.~\ref{fig:scatter}.
Plots for the three other cases are nearly identical.
We classify input-output pairs as ``good'' or ``bad'' as follows.
First we choose a confidence level $0\le c \le 1$ ($c=0.7$ for the data
shown in Fig.~\ref{fig:scatter}). A particular input-output pair is considered
to be good if $x^2 \ge c$ and $y^2\ge c$.
In Fig.~\ref{fig:scatter} the good (bad) pairs are shown by black (gray) markers.

At a fairly low confidence level of $c=0.7$, the region of stable
operation of the $G$ operation is rather small. This corroborates
our earlier finding that successive applications of $G$, e.g. $G^3$,
rapidly drive the system into a region of instability.
In general, quantum systems are very sensitive to noise and become
more sensitive as the number of operations on qubits is
increased\cite{Vedral:1998}.
In the absence of dissipation, it is easy for the system to
leave the relatively small manifold of good input states, a characteristic
feature of almost chaotic dynamics.

\subsection{Hard nonselective pulses}
Another physical implementation of a 2-qubit QC employs
the nuclear spins of two $^1$H spin-1/2
nuclei in a solution of cytosine in D$_2$O\cite{Jones:1998,Jones:1998nat,Jones:1999}.
This system can also be described by Hamiltonian~(\ref{eq:HAM}).
In the NMR experiments hard nonselective pulses are used to
address the qubits. In this section we study the stability
of QC operation for this physical realization of a QC.

As usual it is expedient to transform to another frame of reference that
rotates with a constant frequency.
This is accomplished by substituting in the TDSE
\begin{align}
\KET{\Phi(t)}=e^{ith_{z}(g_1+g_2)(S^z_1+S^z_2)/2
}
\KET{\Psi(t)}
.
\end{align}
The time evolution of $\KET{\Psi(t)}$ is then governed by the Hamiltonian
\begin{align}
\label{eq:HAMjones}
H =- J_{z} S_1^z S_2^z - {\Omega\over2} S_1^z + {\Omega\over2} S_2^z
,
\end{align}
where $\Omega=h_z(g_1-g_2)$.
Guided by experiment\cite{Jones:1998,Jones:1999}
in our numerical work we will set $\Omega/2=1$ and $J_z/\pi\Omega=-0.01887$
(in dimensionless units).

We now consider the time evolution of the two spins when we apply a static
magnetic field $h_x$ along the $x$-axis. The Hamiltonian
in the laboratory frame is given by
\begin{align}H =- J_{z} S_1^z S_2^z - h_{z} (g_1 S_1^z + g_2 S_2^z)
- h_{x} (g_1 S_1^x + g_2 S_2^x)
,
\end{align}
and the corresponding expression in the rotating frame of reference
reads
\begin{align}
\label{eq:H1}
H =&- J_{z} S_1^z S_2^z - {\Omega\over2} S_1^z + {\Omega\over2} S_2^z \nonumber \\
&- h_{x} (g_1 S_1^x + g_2 S_2^x) \cos\Omega t - h_{x} (g_1 S_1^y + g_2 S_2^y) \sin\Omega t.
\end{align}
If the duration of the pulse is short, i.e. $\Omega t\ll1$,
it is a good approximation to drop the time-dependence in~(\ref{eq:H1})
and we obtain
\begin{align}
\label{eq:Htilde}
\widetilde H = - {\Omega\over2} S_1^z + {\Omega\over2} S_2^z - h_{x} (g_1 S_1^x + g_2 S_2^x),
\end{align}
where we used the fact that since $|J_z|\ll|\Omega/2|$,
for short pulses the effect of the spin-spin interaction
on the time evolution is small and may be neglected.
It is instructive to compute the time evolution of the spins,
initially in state $\KET{0}$ ($\KET{0}=\KET{\uparrow}$ by convention),
under these circumstances.
A straightforward calculation yields
\begin{align}
\label{eq:Qj}
Q_j=
\BRACKET{0}{e^{it\widetilde H}S_j^z e^{-it\widetilde H}|0}
={1\over2}-{[2h_xg_j(\Omega/2+\lambda)]^2\over
[(\Omega/2+\lambda_j)^2+h_x^2g_j^2]^2}\sin^2 {\lambda_j t\over2}
,
\end{align}
where $\lambda_j=(\Omega^2/4+h_x^2g_j^2)^{1/2}$.
Expression~(\ref{eq:Qj}) shows that for hard pulses ($|h_x g_j|\gg\Omega/2$)
and $\lambda_j t=\pi$, the effect of the pulse is to change
qubit $j$ from $\KET{0}$ to approximately $\KET{1}$.
Note however that a sequence of such pulses can never
turn $\KET{0}$ into $\KET{1}$ exactly and that both spins are affected
by the pulse.

In the rotating frame of reference the two spins
rotate around the $z$-axis in the opposite sense. This fact
can be exploited to perform operations that leave one spin intact
while flipping the other spin \cite{Jones:1999,Ernst:1987,Freeman:1996}.
For instance, to rotate spin 1 around
90 degrees (clockwise) about the $x$-axis and without changing
the state of spin 2, an operation we call $X_1$,
we can use the sequence of pulses
\begin{align}
\label{eq:Xfromprimes}
{\NOBAR X_1^{\phantom{\prime}}}=
{\NOBAR X_{1,2}^\prime}
{\BAR Y_{1,2}^{\phantom{\prime}}}
{\NOBAR Z_{1,2}^\prime}
{\NOBAR Y_{1,2}^{\phantom{\prime}}}
,
\end{align}
where
${\NOBAR X_{1,2}^\prime}$ denotes a 45-degree pulse around the $x$-axis
acting simultaneously on both spins (see~(\ref{eq:Htilde})),
${\NOBAR Y_{1,2}}$ a 90-degree pulse around the $y$-axis, and
${\NOBAR Z_{1,2}^\prime}$ represents a pulse during which
the spins evolve in time according to~(\ref{eq:Htilde}) and make a
rotation of 45 degrees around the $z$-axis,
(clockwise in the case of spin 1, anti-clockwise in the case of spin 2).
There are many sequences of two-spin pulses that yield $X_1$ (or, as a matter
of fact, any other rotation). For example the inverse of
${\NOBAR X_1^{\phantom{\prime}}}$,
${\BAR X_1^{\phantom{\prime}}}$, can be obtained directly from~(\ref{eq:Xfromprimes}), i.e.
\begin{align}
{\BAR X_1^{\phantom{\prime}}}=
{\BAR Y_{1,2}^{\phantom{\prime}}}
{\BAR Z_{1,2}^\prime}
{\NOBAR Y_{1,2}^{\phantom{\prime}}}
{\BAR X_{1,2}^\prime}
,
\end{align}
or can also be written as
\begin{align}
{\BAR X_1^{\phantom{\prime}}}=
{\BAR X_{1,2}^\prime}
{\BAR Y_{1,2}^{\phantom{\prime}}}
{\BAR Z_{1,2}^\prime}
{\NOBAR Y_{1,2}^{\phantom{\prime}}}
.
\end{align}
For our numerical calculations we have chosen to work with
the sequences 
\begin{subequations} 
\begin{align} 
{\NOBAR X_1^{\phantom{\prime}}}&= {\NOBAR X_{1,2}^\prime} {\BAR Y_{1,2}^{\phantom{\prime}}} 
{\NOBAR Z_{1,2}^\prime}{\NOBAR Y_{1,2}^{\phantom{\prime}}} 
, \\
{\BAR X_1^{\phantom{\prime}}}&= {\BAR X_{1,2}^\prime}{\BAR Y_{1,2}^{\phantom{\prime}}}
{\BAR Z_{1,2}^\prime}{\NOBAR Y_{1,2}^{\phantom{\prime}}}
,\\
{\NOBAR X_2^{\phantom{\prime}}}&={\NOBAR X_{1,2}^\prime}{\BAR Y_{1,2}^{\phantom{\prime}}}
{\BAR Z_{1,2}^\prime}{\NOBAR Y_{1,2}^{\phantom{\prime}}}
, \\
{\BAR X_2^{\phantom{\prime}}}&={\BAR X_{1,2}^\prime}{\BAR Y_{1,2}^{\phantom{\prime}}}
{\NOBAR Z_{1,2}^\prime}{\NOBAR Y_{1,2}^{\phantom{\prime}}}
, \\
{\NOBAR Y_1^{\phantom{\prime}}}&= {\NOBAR Y_{1,2}^\prime}
{\BAR X_{1,2}^{\phantom{\prime}}} {\NOBAR Z_{1,2}^\prime}{\NOBAR X_{1,2}^{\phantom{\prime}}}
, \\
{\BAR Y_1^{\phantom{\prime}}}&= {\BAR Y_{1,2}^\prime}
{\BAR X_{1,2}^{\phantom{\prime}}} {\BAR Z_{1,2}^\prime}{\NOBAR X_{1,2}^{\phantom{\prime}}}
, \\
{\NOBAR Y_2^{\phantom{\prime}}}&={\NOBAR Y_{1,2}^\prime}{\BAR X_{1,2}^{\phantom{\prime}}}
{\BAR Z_{1,2}^\prime}{\NOBAR X_{1,2}^{\phantom{\prime}}}
, \\
{\BAR Y_2^{\phantom{\prime}}}&={\BAR Y_{1,2}^\prime}{\BAR X_{1,2}^{\phantom{\prime}}}
{\NOBAR Z_{1,2}^\prime}{\NOBAR X_{1,2}^{\phantom{\prime}}}
, 
\end{align}
\end{subequations} 
to perform the single-qubit operations that are used in the
implementation of Grover's search algorithm.
As in the case of resonant pulses
the natural time-evolution of the system provides the tool to
carry out the conditional phase-shift operation $I(\pi)$.

The numerical values for
the qubits 1 and 2 in the final state of the machine, obtained
by running the QCE for this particular
hardware realization of a QC and QA are 
are given in Table~\ref{tab:grover2}.

From a comparison with the data of Table~\ref{tab:grover} we conclude
that this hardware realization of a 2-qubit QC seems to be
much less sensitive to QPP errors.
As a matter of fact, in our numerical experiments (data not
shown but included in the QCE download) we found none.
The main reason for the good performance of this QC can be traced back
to the fact that the QC contains two spins with exactly
the same precession frequency, precessing in opposite direction.
It is clear that it will become rather cumbersome if not impossible to
design a $N$-qubit QC (if $N\gg2$) using this approach:
We would have to solve a complicated optimization problem
that for each QA we would like to run on the QC.

\subsection{Reducing systematic errors}
For QC hardware to be of practical use,
a basic requirement is that logically identical but physically
different implementations of the same QA yield the same answers.
The examples discussed above show that optimization of the pulses
is essential to achieve this goal.
For 2-qubit QC's, the optimization
problem is relatively easy to solve because only two frequencies are involved.
Alternatively one may choose to use the same duration for all pulses and
optimize the strengths of the pulses.
However, in the case of $N\gg2$ qubits,
choosing pulse durations commensurate with $N$ different frequencies
will considerably slow down the operation of the whole machine.
In addition, the longer the pulses, the more important the effect
of imperfections of the pulses becomes. Although there
are ingenious techniques to optimize pulses in this 
respect\cite{Ernst:1987,Tseng:2000,Doria:2000,Leung:2000},
it may well be that in order to successfully run a QA on
$N$-qubit QC hardware
one first has to solve a rather complicated optimization problem
and then simulate the QC by running the QA on a conventional machine.

\section{Can dissipation reduce the QPP ?}
\label{sec:dissip}
Elsewhere we reported on the effect of dissipation on the quantum
dynamics of nanoscale magnets\cite{Saito:1999,Miyashita:2000}.
In these systems dissipation usually causes decoherence.
Decoherence limits the time for performing quantum computations.
Above we have demonstrated that programming non-ideal QC's is
difficult due to intrinsic instabilities of the physical device.
Therefore we wonder if dissipation can provide the stabilizing
processes, perhaps at the cost of reducing the time of
coherent quantum operation.

Let us consider the procedure to generate the uniform superposition state
from the state with all spins up.
Above we made use of sequences such as $XXY$ but it
is easy to see that from a programming-point-of-view
$Y$ is just as good: In a classical picture both sequences
change the direction of a spin from the $z$ to the $x$-axis.
Thus, on an ideal QC we may omit $XX$ if we want.
However, in an experiment (in which there always is
some amount of dissipation), $XX$ can help to stabilize the direction
of the spin once it points in the $x$-direction.
Although in the pure quantum case $XX$
only causes the spin to precess around the $x$-axis,
the damping of this motion by dissipation
will let the spin relax to the $x$-direction.

We have studied the effects of dissipation on quantum computations
using the master equation\cite{Kubo:1985}
\begin{align}
\label{eq:master}
{\partial \rho(t)\over \partial t}=
-i
[H,\rho(t)] -\lambda \left(
[C,R\rho(t)]+[C,R\rho(t)]^\dagger\right)
,
\end{align}
that describes the time-evolution of
the reduced (to the subspace of the qubits) density matrix $\rho(t)$.
The operator $R$ is defined as
\begin{align}
\BRACKET{k}{R|m}=\zeta(E_k-E_m)N_\beta(E_k-E_m)\BRACKET{k}{C|m}
,
\end{align}
where $H\KET{k}=E_k\KET{k}$, $H$ being the Hamiltonian describing
the QC. We take for the spectral
density of the bosonic thermal bath
$\zeta(E)=I_0 E^2 \hbox{sign}(E)$, i.e. an super-Ohmic reservoir,
and $N_\beta(E)=(e^{\beta E}-1)^{-1}$ denotes the boson occupation number\cite{Saito:2000}.
The operator $C$ specifies the coupling between the spins and the
bosons, e.g. $C=\sum_j (S_j^x+S_j^z)/2,$.
We set the inverse temperature $\beta=100$\cite{NMRtemp}.
The parameter $\lambda$ controls the flow of heat
to and from the QC.

The procedure for simulating a QC in the presence of dissipation
is very similar to the one used in the pure quantum case:
The only difference is
that we solve the master equation~(\ref{eq:master}) instead of TDSE~(\ref{eq:TDSE}).
For $\lambda=0$ the two completely different numerical methods used to solve
these two equations (see \cite{Saito:2000} and \cite{DeRaedt:2000qce}
respectively) yield identical results.

\begin{figure}[t]
\def\epsfsize#1#2{1.0#1} 
\begin{center}
\epsffile{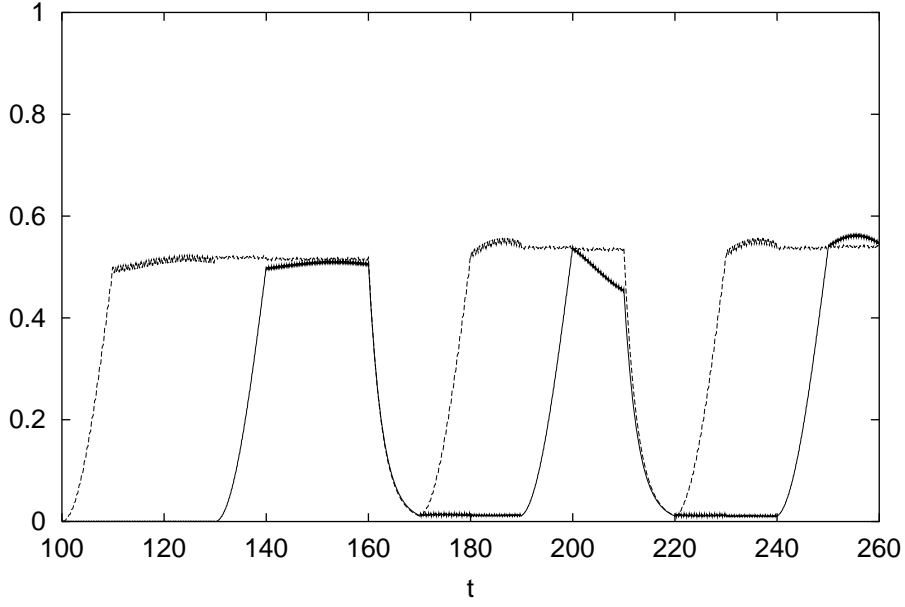}
\end{center}
\caption{
Time evolution of the qubits $\tilde Q_1^\prime$ (solid line) and
$\tilde Q_2^\prime$ (dashed line) obtained by executing $U_2\KET{u^\prime}$
on the physical model of an NMR-like QC,
in the presence of dissipation ($\lambda=10^{-5}$).
The time intervals for each operation have been
rescaled to make them look equal.}
\label{fig:dissipation}
\end{figure}

We have studied the effect of dissipation
by performing calculations for the same cases as described in Section~\ref{sec:respulses}.
Our simple-minded model of dissipation is found to cause only decoherence.
We don't find qualitative differences between the time-evolutions
corresponding to the two different initial conditions.
The presence of dissipation does not seem to affect the QPP.
For $\lambda=10^{-5}$
the value of the qubits in the final states is approximately 1/2,
i.e. all useful information is lost (see Fig.~\ref{fig:dissipation}).
Apparently our choice does not lead to a relaxation during the $XX$ pulses.
Although the dissipation processes incorporated through the
use of master equation~(\ref{eq:master}) lead to the correct
thermal equilibrium state, there is no unique prescription
to choose the bath Hamiltonian and the coupling between
the bath and the spin system, i.e. the form of $C$.
We are currently studying more realistic mechanisms
for dissipation and will report on the results elsewhere.

\section{Four-qubit QC's}
\label{sec:fourqubits}
In general practical applications of QC's will require complicated
QA's. Therefore it is of interest to study how the instabilities
due to imperfections in the operations affect the performance
in such complicated QA's. In this section we use a copy (or swap)
procedure in a four-qubit QC as one of the simplest examples.

As discussed above, an implementation of Grover's search algorithm
on a QC consist of two separate parts: 1) The preparation of
the uniform superposition state $\KET{u}$ and encoding of the database
information and 2) the search (query) of the database.
The algorithms described above use the same two qubits for
these two different tasks.
First they are used to store the information
contained in the database and second they are used to carry out
Grover's algorithm to query the very same two qubits.
Although this is sufficient to demonstrate the realization of 2-qubit
QC hardware, from the point of view of computation this demonstration
is of little use. Indeed, instead of organizing the computation
in two steps (bringing the two qubits into a state that reflects
the content of the database and then using the same two
qubits to perform the query) we could have written the information
from the database directly into the qubits in a form which reveals
the position of the item in a trivial manner.

\begin{figure}[t]
\def\epsfsize#1#2{.80#1} 
\begin{center}
\epsffile{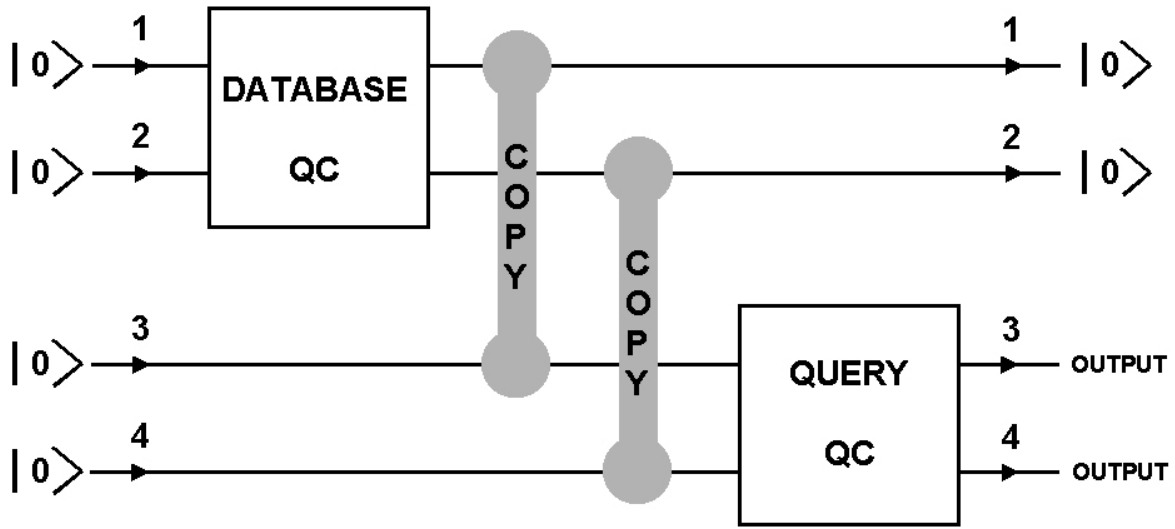}
\end{center}
\caption{
Structure of a quantum program for Grover's search algorithm,
using a 2-qubit QC to hold the database information and
another 2-qubit QC to perform the search.}
\label{fig:4qubits}
\end{figure}

A computationally non-trivial demonstration of a QC running Grover's
database search algorithm requires the database and quantum processor
to have their own qubits.
This obviously means using
at least four qubits, two to hold the database information (in superposition
state) and two to process the query.
Therefore we will need an operation to copy the state of the
database into the quantum processor.
A schematic picture of how the calculation is organized is shown in
Fig.~\ref{fig:4qubits}.
Let us now study how instabilities in the physical operations affect
the outcome of this more complicated computational procedure.

The copy operation transfers the amplitudes of
an arbitrary linear combination of spin-up and spin-down of e.g. spin 1
to e.g. spin 2, initially in a state of spin up. More specifically
\begin{align}
\label{eq:copydef}
C_{1,2}(a\KET{0}_1+b\KET{1}_1)\KET{0}_2
=
\KET{0}_1(a\KET{0}_2+b\KET{1}_2)
.
\end{align}
In principle this can be done by a network of CNOT gates\cite{Ekert:1996}.
One possible realization of a CNOT gate is the pulse sequence
\begin{align}
C_{1,2}={\NOBAR Y}_1 {\NOBAR X}_1
{\NOBAR Y}_1 {\NOBAR X}_1 {\BAR Y}_1 {\NOBAR X}_2 {\BAR Y}_2
I(\pi)
{\NOBAR Y}_2
,
\end{align}
up to irrelevant phase factors, but there are many others.
A fundamentally different sequence that performs the same task is
\begin{align}
\label{eq:copyxy}
C_{1,2}=
{\NOBAR Y}_2 {\BAR X}_2 {\BAR Y}_2 I^\prime(\pi)
,
\end{align}
where
\begin{align}
I^\prime(2a)
\left(
  \begin{array}{l}
\KET{00}\\ \KET{10}\\ \KET{01}\\ \KET{11}
  \end{array}
\right)
\equiv
\left(
  \begin{array}{cccc}
1&0& 0 &0\\
0&\cos a& i\sin a &0\\
0&i\sin a& \cos a& 0\\
0&0&0&1
\end{array}
\right)
\left(
  \begin{array}{l}
\KET{00}\\ \KET{10}\\ \KET{01}\\ \KET{11}
  \end{array}
\right)
.
\end{align}

A simple calculation shows that
the two-qubit operation $I^\prime(2a)$ corresponds
to the time-evolution of a two-spin XY model:
\begin{align}
\label{eq:Ixy}
I^\prime(2\tau J_{xy})=e^{-i\tau J_{xy} (S_1^x S_2^x + S_1^y S_2^y)}
.
\end{align}

The other pulses in sequence~(\ref{eq:copyxy}) serve to remove unwanted phase factors.
Note that~(\ref{eq:copydef})
 and~(\ref{eq:Ixy}) destroy the state of the database,
a manifestation of the ``no-cloning theorem''\cite{DAriano:1996}. 
As Fig.~\ref{fig:4qubits} shows this is not really a problem as the state
of the database after the copy operation took place is
the same state that was used to initialize the state of the database.

As before our aim is to address the fidelity of the results
obtained by running a quantum algorithm on a hardware realization of a QC.
Our QCE can simulate various
candidate technologies and architectures (ways of interconnecting
different units). However, to analyse the QPP
it is at present sufficient to consider marginal extensions
of two-qubit QC's.
With this in mind we have chosen to
``implement'' a 4-qubit QC as two identical Ising models
\begin{align}
\label{eq:HAMfour}
H =- J_{z} S_1^z S_2^z - h_{z} (g_1 S_1^z+ g_2 S_2^z)
   - J_{z} S_3^z S_4^z - h_{z} (g_1 S_3^z+ g_2 S_4^z)
.
\end{align}

In our numerical work we take for the values of the model parameters
those corresponding to the chloroform molecule and we use selective
resonant pulses to address the individual spins (see above).
Thereby we assume, again for simplicity of analysis, that we
can address pairs (1,2) and (3,4) separately (i.e. a pulse tuned
to spin 1 does not affect spin 3 etc.).
In the case where we use CNOT's to perform the copy-qubit
operation we add to~(\ref{eq:HAMfour}) the Ising interactions
\begin{subequations} 
\begin{align}
H_{1,3} &=- J_{z} S_1^z S_3^z, \\
H_{2,4} &=- J_{z} S_2^z S_4^z,
\end{align}
\end{subequations} 
or, in the case the XY-model is used to
copy the qubits from the database into the query processor,
we add to~(\ref{eq:HAMfour})
\begin{subequations} 
\begin{align}
H_{1,3}^\prime &=- J_{xy} (S_1^x S_3^x + S_1^y S_3^y), \\
H_{2,4}^\prime &=- J_{xy} (S_2^x S_4^x + S_2^y S_4^y).
\end{align}
\end{subequations} 
It is obvious that these operations cannot be realized in terms
of NMR technology.
However this extension from a 2-qubit to a 4-qubit QC is
sufficient to study in detail the stability of QA's.

\subsection{Resonant pulses}
In our numerical simulations on the QCE we take the same model parameters
as those of the chloroform molecule (see above) for both
the database and query machine.
We use resonant pulses to address the single qubits.
Thereby we assume that qubits 1 and 2 can be shielded
from qubits 3 and 4 during the application of these pulses.

The results of running Grover's search algorithm on
the QCE are 
summarized in Table~\ref{tab:4bitgrover}.
In these calculations we used the pulses optimized in the sense
discussed above (i.e.\ when used to run on a 2-qubit QC they
yield the correct answers, independent of the initialization sequence).
Nevertheless we observe that the implementation that uses
the CNOT gates performs rather poorly, at the
border of giving the wrong answers.
As in the 2-qubit case, the fact that the duration of the pulses
is no longer commensurate with the Larmor periods of the spins
is one source of errors. Another mechanism that
causes errors to accumulate is discussed below.

\subsection{Rotating field}
Finally, in an attempt to further reduce the uncertainty on the outcome
of the quantum computations,
we have repeated the simulations using resonant pulses
of rotating fields\cite{Slichter:1990,Baym:1974}.
More specifically, to apply a pulse to rotate for example spin 1,
we add to the Hamiltonian a term of the form
\begin{align}
H_1(t) = h_{1} S_j^y \sin f_{1} t +h_{1}S_j^x \cos f_{1} t,
\end{align}
where the frequency $f_1$ is tuned to the resonance frequency
of spin 1 and the intensity of the pulse $h_{1}$ is determined
by the desired angle of rotation.
A simple calculation shows that such a pulse
transforms the single qubit in {\sl exactly}
the same manner as the corresponding, ideal transformation\cite{Slichter:1990,Baym:1974}.
It is evident that this kind of optimization should
improve the stability of the QA with respect to
logically-allowed interchanges of operations.
Further improvements can be made by reordering
some of the elementary operations
(recall that the sequence implementing e.g. a CNOT is not unique).
The results of the simulations are presented in 
Table~\ref{tab:4bitgroverrot}.
Compared to the results of Table~\ref{tab:4bitgrover} there is one major change:
The CNOT based implementation performs much better.

In this implementation there are two sources of
errors: The effect of the pulses on non-resonant spins and
the interaction-controlled phase shift.
In the case of the chloroform molecule
there is a large difference in time scale (roughly a factor $10^6$)
between the precession frequencies and the spin-spin
interaction. Therefore, during the execution of
an interaction-controlled phase shift, the spins make
a large number of rotations about the $z$-axis. We have seen
that it really matters for the stability
of the calculation whether the angles of rotation
are a multiple of $2\pi$ or not.
However, in view of the difference in time scale, this implies
that the duration of an interaction-controlled phase shift must be
specified with a sufficiently high accuracy in order
to recover the correct final results.

\section{Summary}
We have studied the difficulties that are encountered in programming
quantum computer hardware.
Taking Grover's database quantum search algorithm as an example,
we ran various, logically identical versions of the algorithm
on different, physically realizable quantum computer hardware.
We demonstrated that the choice of the physical processes used
for quantum computation has direct consequences
in terms of programming the machine.

We have shown that in the absence of dissipation
non-ideal, physically realizable quantum computers
operate in a regime of extreme sensitivity.
This high sensitivity reflects itself in problems of programming
quantum computers such that they perform correct calculations.
These problems cannot be solved by error correction but
may be reduced, and in some cases almost eliminated,
at the cost of extensive, machine and program-specific optimization.
A possible route to reduce this sensitivity may be to
introduce some form of dissipation in a well-controlled manner.
Dissipation can extend the region of stable operation
of the quantum computer but also limits
the time interval for quantum computer operation due to decoherence.
We are currently exploring ideas along this line.

\section*{Acknowledgement}
Support from the Dutch NCF, FOM and NWO,
and from the Grant-in-Aid for Research from the
Japanese Ministry of Education, Science and Culture
is gratefully acknowledged.

\appendix

\section{Quantum Computer Emulator (QCE)} \label{app:qce}

\begin{figure}[t]
\def\epsfsize#1#2{.80#1} 
\begin{center}
\epsffile{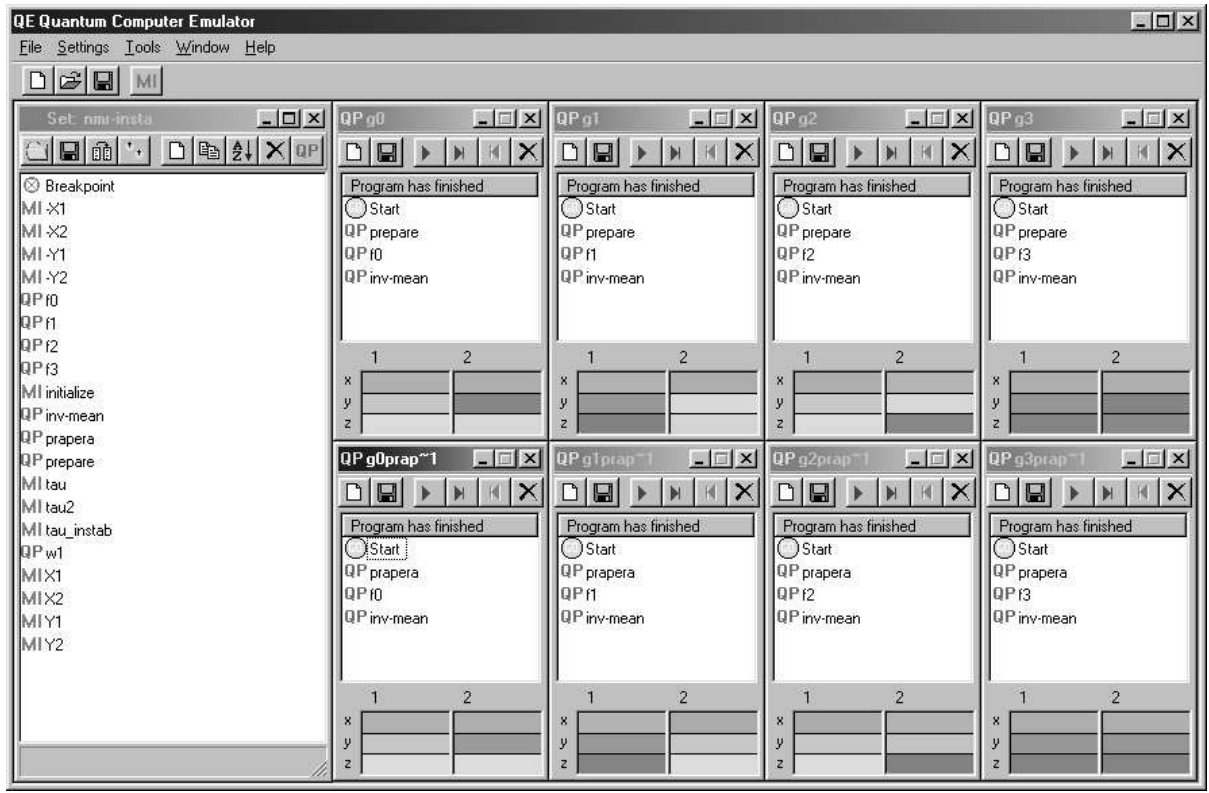}
\end{center}
\caption{Snapshot of the Quantum Computer Emulator showing a window with a
set of operations implementing a
quantum computer using resonant pulses and
windows with quantum programs implementing Grover's
database search.
}
\label{fig:qcesnap}
\end{figure}

In this appendix, we show an example of QCE. 
In Fig.~\ref{fig:qcesnap} we show a snapshot of a window of QCE
with a set of operations used to produce
the results of Table~\ref{tab:grover} for the optimized pulses.

The left panel in Fig.~\ref{fig:qcesnap} lists the fundamental operations,
where MI stands for micro instruction. For example MI `-X1'
means the operation ${\bar X_1}$, etc.
MI's are defined by a table with interaction parameters, duration time, etc.
This table appears when we double-click an MI.
More details can be found in \cite{DeRaedt:2000qce}.
The set of MI's for the case of resonant optimized pulses is stored
under the name `nmr-insta'.
A QP represents a subroutine which consist of sequence of MI's.
For example QP `f2' means the sequence of operations $F_2$, see Eq.~(\ref{eq:f2}).
The MI `initialize' simply prepares the state $\KET{00}$,
MI `inv-means' performs the inversion about the mean (the essential
step in Grover's algorithm, see Eq.~(\ref{eq:Gmat})),
MI `tau' corresponds to the operation $I(\pi)$ (see Eq.~(\ref{eq:Ia})), QP `prepare' denotes the
preparation of $\KET{u}$ while QP `prapera' denotes the preparation of
$\KET{u^\prime}$, etc.
The MI's `tau2', `tau\_insta', and `w1' are not used in this example.

The eight small windows show the programs corresponding to $U_0,\cdots U_3$
using $\KET{u}$ (upper level) and those using $\KET{u^\prime}$ (lower level),
which are named QP `g0' to QP `g3prap\~1'.
The result of running each program is shown in the grid below the QP.
The expectation value of $(1-S_i^{\alpha})/2$, $i=1$ or 2, and $\alpha=x,y$ or z,
can be read off from a particular cell by moving the mouse over the cell (not shown).
In Fig.~\ref{fig:qcesnap}
the approximate value can be determined from the darkness of the cell:
Light gray corresponds to zero, dark gray to one.
For example, the value of the cell $(z,1)$ in QP `g0' is 0.027 and that of $(z,2)$
is 0.152, as given in Table~\ref{tab:grover} ($Q_1'$ and $Q_2'$ respectively).
For other cases, e.g. the non-optimized pulses, we use another set
of MI's (`nmr') with parameters given in \cite{DeRaedt:2000qce}.

\bibliography{abbrevs,lz,qc,trotter,spinsystems,other,notes}

\begin{thebibliography}{10}
\bibitem{NMRtemp}
In terms of the NMR experiments referred to in this paper, this is an extremely
  low temperature.
\end{thebibliography}
\bibliographystyle{prsty}        

\begin{table}[ht]
  \caption{Final state of the qubits after running
the Grover's database search algorithm on an
ideal QC ($Q_1,Q_2$), on a NMR-like QC ($\hat Q_1,\hat Q_2$),
and on the same QC ($\tilde Q_1,\tilde Q_2$)
using a different, but logically identical, initialization sequence
($\KET{\BAR u}=\KET{u^\prime}$ instead of $\KET{\BAR u}=\KET{u}$).
The optimized results, marked with a prime, have been obtained by changing
the duration and intensity of the pulses tuned to the resonance frequency
of spin 1 by a factor of two and 1/2 respectively.}
  \begin{center}
    \begin{tabular}{ccccc}
      &$U_0 \KET{\BAR u}$&$U_1\KET{\BAR u}$& $U_2\KET{\BAR u}$&$U_3\KET{\BAR u}$ \\
\hline
$Q_1$ &  0      & 1     & 0     & 1  \\
$Q_2$     &  0      & 0     & 1     & 1  \\
\hline
$\hat Q_1$&  0.028  & 0.966 & 0.037 & 0.995  \\
$\hat Q_2$&  0.163  & 0.171 & 0.836 & 0.830  \\
\hline
$\tilde Q_1$& 0.955 & 0.041 & 0.971 & 0.027 \\
$\tilde Q_2$& 0.031 & 0.026 & 0.971 & 0.972  \\
\hline
$\hat Q_1^\prime$&  0.027  & 0.972 & 0.037 & 0.964  \\
$\hat Q_2^\prime$&  0.152  & 0.180 & 0.847 & 0.820  \\
\hline
$\tilde Q_1^\prime$& 0.030 & 0.969 & 0.034 & 0.965 \\
$\tilde Q_2^\prime$& 0.022 & 0.035 & 0.977 & 0.965 
    \end{tabular}
    \label{tab:grover}
  \end{center}
\end{table}

\begin{table}[ht]
    \caption{Final state of the qubits after running the query part $G$ of
Grover's database search algorithm on an
ideal QC ($Q_1,Q_2$) and on a NMR-like QC ($Q_1^\prime,Q_2^\prime$).
In all cases the input state $\KET{\Psi_2}$ corresponds exactly
to the case where the item is located in position 2 in the database.
The queries ${\hat G}$ and ${\tilde G}$ are logically identical to $G$.
The differences in the outputs of the NMR-like QC are
due to the internal quantum dynamics of the physical qubits used.}
  \begin{center}
    \begin{tabular}{ccccccc}
& $G\KET{\Psi_2}$& ${\hat G}\KET{\Psi_2}$& ${\tilde G}\KET{\Psi_2}$&
$G^3\KET{\Psi_2}$&${\hat G}^3\KET{\Psi_2}$ &${\tilde G}^3\KET{\Psi_2}$  \\
\hline
$Q_1$     &  0      & 0     & 0     & 0 & 0     & 0 \\
$Q_2$     &  1      & 1     & 1     & 1 & 1     & 1 \\
\hline
$Q_1^\prime$&  0.257  & 0.154 & 0.487 & 0.908 & 0.766 & 0.578 \\
$Q_2^\prime$&  0.944  & 0.944 & 0.988 & 0.938 & 0.857 & 0.995 
    \end{tabular}
    \label{tab:G}
  \end{center}
\end{table}

\begin{table}[ht]
    \caption{Final state of the QC after running
the Grover's database search algorithm on an
ideal QC ($Q_1,Q_2$) and an QC ($\hat Q_1,\hat Q_2$)
using hard, non-selective pulses.
($\tilde Q_1,\tilde Q_2$): QC using hard non-selective pulses,
and a different but logically identical initialization sequence
($\KET{\BAR u}=\KET{u^\prime}$ instead of $\KET{\BAR u}=\KET{u}$).}
  \begin{center}
    \begin{tabular}{ccccc}
&$U_0\KET{\BAR u}$&$U_1\KET{\BAR u}$& $U_2\KET{\BAR u}$&$U_3\KET{\BAR u}$ \\
\hline
$Q_1$     &  0      & 1     & 0     & 1 \\
$Q_2$     &  0      & 0     & 1     & 1 \\
\hline
$\hat Q_1$&  0.000  & 0.999 & 0.001 & 0.999 \\
$\hat Q_2$&  0.000  & 0.005 & 0.999 & 0.996 \\
\hline
$\tilde Q_1$&  0.001  & 0.999 & 0.002 & 0.999 \\
$\tilde Q_2$&  0.001  & 0.002 & 0.999 & 0.998 
    \end{tabular}
    \label{tab:grover2}
  \end{center}
\end{table}

\begin{table}[ht]
    \caption{Final state of the QC after running
the Grover's database search algorithm on a 4-qubit QC.
$Q_3,Q_4$: ideal QC;
$\hat Q_3,\hat Q_4$: NMR-like QC, using
an XY-model time-evolution to copy the state of the database
to the two-qubit QC that performs the query;
$\tilde Q_3,\tilde Q_4$: NMR-like QC, using
a sequence of CNOT to perform the copy operations.
The values of the two database qubits are not shown.}
  \begin{center}
    \begin{tabular}{ccccc}
& $U_0\KET{u}$ & $U_1\KET{u}$& $U_2\KET{u}$&$U_3\KET{u}$ \\
\hline
$Q_3$     &  0      & 1     & 0     & 1 \\
$Q_4$     &  0      & 0     & 1     & 1 \\
\hline
$\hat Q_3$& 0.172  & 0.809 & 0.190 & 0.831  \\
$\hat Q_4$& 0.135  & 0.187 & 0.865 & 0.814  \\
\hline
$\tilde Q_3$&  0.412  & 0.565 & 0.433 & 0.591 \\
$\tilde Q_4$&  0.113  & 0.218 & 0.888 & 0.783 
    \end{tabular}
    \label{tab:4bitgrover}
  \end{center}
\end{table}

\begin{table}[ht]
    \caption{Final state of the QC after running
the Grover's database search algorithm on a 4-qubit QC,
using pulses optimized to reduce the effect
of imperfections of the operations and phase errors.
$Q_3,Q_4$: ideal QC;
$\hat Q_3,\hat Q_4$: NMR-like QC, using
an XY-model time-evolution to copy the state of the database
to the two-qubit QC that performs the query;
$\tilde Q_3,\tilde Q_4$: NMR-like QC, using
a sequence of CNOT to perform the copy operations.
The values of the two database qubits are not shown.}
  \begin{center}
    \begin{tabular}{ccccc}
&$U_0\KET{u}$&$U_1\KET{u}$& $U_2\KET{u}$&$U_3\KET{u}$ \\
\hline
$Q_3$     &  0      & 1     & 0     & 1 \\
$Q_4$     &  0      & 0     & 1     & 1 \\
\hline
$\hat Q_3$ & 0.195  & 0.806 & 0.193 & 0.806 \\
$\hat Q_4$ & 0.195  & 0.194 & 0.805 & 0.806 \\
\hline
$\tilde Q_3$ & 0.265  & 0.737 & 0.262 & 0.737 \\
$\tilde Q_4$ & 0.197  & 0.193 & 0.804 & 0.807 
    \end{tabular}
    \label{tab:4bitgroverrot}
  \end{center}
\end{table}

\newpage

\newpage

\newpage

\newpage

\end{document}